\documentclass[runningheads]{llncs}

\usepackage[T1]{fontenc}
\usepackage{amsmath}
\usepackage{booktabs}
\usepackage{graphicx}
\usepackage{hyperref}
\usepackage{booktabs}
\usepackage{multirow}
\usepackage{pifont}
\usepackage{fontawesome5}
\usepackage{xcolor}
\usepackage{subcaption}
\usepackage{amssymb}
\usepackage{array}
\usepackage[most]{tcolorbox}
\usepackage{amsmath,amsfonts,bm}
\usepackage{float}

\usepackage{xr-hyper}
\newif\ifdraft

\ifdraft

\newcommand{\KK}[1]{{\color{blue}{\bf KK: #1}}}

\else
\newcommand{\KK}[1]{}

\fi

\newcolumntype{Y}{>{\raggedright\arraybackslash}X}

\definecolor{promptpurple}{RGB}{242, 242, 255}
\definecolor{promptgreen}{RGB}{242, 255, 242}
\definecolor{promptblue}{RGB}{242, 248, 255}
\definecolor{skillback}{RGB}{255, 252, 242}
\definecolor{bordergray}{RGB}{80, 80, 80}
\newtcolorbox{promptbox}[1]{
  breakable,
  enhanced,
  colback=#1,
  colframe=bordergray,
  boxrule=1pt,
  sharp corners,
  left=15pt,
  right=15pt,
  top=12pt,
  bottom=12pt,
  before skip=6pt,
  after skip=10pt,
  fontupper=\small\ttfamily
}
\newtcolorbox{skillbox}{
  breakable,
  enhanced,
  colback=skillback,
  colframe=bordergray,
  boxrule=1pt,
  sharp corners,
  left=15pt,
  right=15pt,
  top=12pt,
  bottom=12pt,
  before skip=6pt,
  after skip=10pt,
  fontupper=\small\ttfamily
}

\definecolor{trajthink}{RGB}{246, 246, 244}
\definecolor{trajthinkframe}{RGB}{74, 78, 82}
\definecolor{trajtool}{RGB}{247, 250, 248}
\definecolor{trajtoolframe}{RGB}{70, 110, 104}
\definecolor{trajerror}{RGB}{255, 244, 235}
\definecolor{trajfinal}{RGB}{235, 250, 240}
\definecolor{trajframe}{RGB}{70, 88, 94}
\definecolor{trajgreen}{RGB}{60, 140, 90}

\newtcolorbox{thinkstep}[1]{
  enhanced, breakable,
  colback=trajthink, colframe=trajthinkframe, boxrule=0.5pt, arc=1pt,
  left=6pt, right=6pt, top=3pt, bottom=3pt,
  before skip=4pt, after skip=4pt,
  title={\faLightbulb~#1}, coltitle=white, colbacktitle=trajthinkframe,
  fonttitle=\bfseries\footnotesize\sffamily,
  fontupper=\small
}

\newtcolorbox{toolmeta}[1]{
  enhanced, breakable,
  colback=trajtool, colframe=trajtoolframe, boxrule=0.5pt, arc=1pt,
  left=6pt, right=6pt, top=3pt, bottom=3pt,
  before skip=4pt, after skip=4pt,
  title={\faWrench~#1}, coltitle=white, colbacktitle=trajtoolframe,
  fonttitle=\bfseries\footnotesize\sffamily,
  fontupper=\small
}

\newtcolorbox{toolstep}[2]{
  enhanced, breakable,
  sidebyside, sidebyside align=top, righthand width=#2,
  colback=trajtool, colframe=trajtoolframe, boxrule=0.5pt, arc=1pt,
  left=6pt, right=6pt, top=3pt, bottom=3pt,
  before skip=4pt, after skip=4pt,
  title={\faWrench~#1}, coltitle=white, colbacktitle=trajtoolframe,
  fonttitle=\bfseries\footnotesize\sffamily,
  fontupper=\small
}

\newtcolorbox{errorstep}[1]{
  enhanced, breakable,
  colback=trajerror, colframe=trajframe, boxrule=0.5pt, arc=1pt,
  left=6pt, right=6pt, top=3pt, bottom=3pt,
  before skip=4pt, after skip=4pt,
  title={\ding{55}~#1}, coltitle=white, colbacktitle=trajframe,
  fonttitle=\bfseries\footnotesize\sffamily,
  fontupper=\small
}

\newtcolorbox{finalstep}[1]{
  enhanced, breakable,
  colback=trajfinal, colframe=trajgreen, boxrule=0.9pt, arc=1pt,
  left=6pt, right=6pt, top=3pt, bottom=3pt,
  before skip=4pt, after skip=6pt,
  title={#1}, coltitle=white, colbacktitle=trajgreen!70!black,
  fonttitle=\bfseries\footnotesize\sffamily,
  fontupper=\small
}

\hypersetup{
  colorlinks=true,
  breaklinks=true,
  linkcolor=blue,
  urlcolor=blue,
  citecolor=blue
}

\begin{document}

\title{MammoClaw: Towards Skill-Evolving Agent Harness for Breast Cancer Mammography Analysis}

\titlerunning{MammoClaw}

\author{Krishna Kanth Nakka}
\authorrunning{K. K. Nakka}

\institute{
Munich, Bavaria, Germany\\
\email{krishkanth.92@gmail.com}\\
}

\maketitle

\begin{abstract}
In this work, we explore \texttt{MammoClaw}, a training-free agent framework that leverages frozen MLLMs for mammography analysis. To support agentic investigation, we equip the agent with lightweight mammography-specific tools for targeted image analysis, including ROI, paired-view, and contralateral-breast examination.
\texttt{MammoClaw} iteratively gathers evidence through these tools, while skill evolution enables non-parametric adaptation by transforming failed trajectories into reusable guidance for later runs. We evaluate the framework on BI-RADS assessment and breast density estimation tasks. In our experiments, we find that tools alone do not reliably improve performance, whereas evolved skills can improve tool-use behavior and performance in some settings. Beyond these results, \texttt{MammoClaw} enables transparent inspection of evidence acquisition, tool interactions, and failure modes, facilitating the analysis and auditing of agent behavior. We view this work as an exploratory study of training-free, self-evolving agentic approaches for mammography and hope it provides a concrete starting point for future work on mammography-specific tools and self-evolution mechanisms. We release our code at \url{https://krishnakanthnakka.github.io/mammoclaw}.
\end{abstract}

\section{Introduction}

Medically fine-tuned multimodal large language models (MLLMs)~\allowbreak\cite{mullappilly2026medix,deria2026medmo,chen2024towards,li2023llava} have demonstrated strong visual reasoning capabilities across diverse medical imaging tasks~\cite{lau2018dataset,liu2021slake,he2020pathvqa}, including mammography~\cite{zhu2025benchmark,ghosh2025mammo,cao2025mammovlm}. This capability, however, is typically obtained through expensive post-training pipelines, including supervised fine-tuning~\cite{radford2018improving,zhou2023lima} and reinforcement learning~\cite{shao2024deepseekmath,ouyang2022training}, which rely on large curated medical datasets, substantial computational resources, and costly expert supervision for reasoning traces, thereby limiting their scalability to new tasks and backbones.

A recent alternative is to investigate whether a frozen, general-purpose MLLM can be enhanced through test-time agentic execution~\cite{wu2026clinseekagent,liu2026physicianbench,zhang2025radagents,roschewitz2026radagent}, by equipping it with domain-specific tools rather than updating its weights. For instance, recent frameworks such as RadAgent~\cite{roschewitz2026radagent} and ClinSeekAgent~\cite{wu2026clinseekagent} suggest that this paradigm can improve performance across several clinical and general radiology domains. Yet, to our knowledge, its application to breast imaging remains largely unexplored. Motivated by this gap, we conduct an exploratory investigation of whether agentic execution with frozen MLLMs can benefit mammography assessment tasks such as BI-RADS grading and breast density estimation.

To this end, we equip a frozen MLLM with a lightweight suite of deterministic, model-free mammography tools for structured evidence acquisition, supporting knowledge retrieval, ROI inspection and paired-view and contralateral-view comparison without relying on costly auxiliary models such as segmentation or detection networks. In practice, however, we find that tool augmentation alone is not sufficient in our evaluation: augmenting the frozen agent with the full tool suite leaves BI-RADS macro F1 essentially unchanged relative to the no-tool agent ($0.108 \rightarrow 0.106$). This observation suggests that the challenge may lie not simply in providing access to tools, but also in how the agent decides when and how to use them. This motivates our further investigation of whether experience from previous failures can provide reusable guidance for evidence acquisition and tool use.

In light of this, we develop and investigate \texttt{MammoClaw}, a training-free agent framework for breast mammography that evolves non-parametrically through reusable skills~\cite{sun2026experience,ma2026skillclaw,yang2026autoskill,xia2026skillrl}. \texttt{MammoClaw} distills reusable guidance from the agent's failed trajectories into skills, which are then retrieved to guide the agent in subsequent rounds. By doing so, the framework provides a testbed for studying whether experience-driven skill evolution can improve reasoning and tool-use behavior without updating the underlying model weights. We evaluate \texttt{MammoClaw} on two mammography tasks, BI-RADS prediction and breast density estimation. Our results suggest that experience-driven skill evolution can provide benefits over the no-tools baseline in our experimental setting. Furthermore, the framework exposes the agent's evidence acquisition and tool interactions, enabling auditing and attribution of failure modes to support trustworthy agentic mammography systems. Overall, we position our work as an initial exploration rather than a comprehensive solution, and use the findings to motivate broader investigation of mammography-specific tools, skill representations, and self-evolving agents. Our contributions are summarized as follows:

\begin{enumerate}
\item \textbf{Framework:} We develop \texttt{MammoClaw}, a training-free agent harness for exploring agentic mammography analysis with frozen MLLMs, deterministic mammography tools, and an automated offline skill-evolution workflow.

\item \textbf{Tool Suite:} We develop an initial suite of lightweight, deterministic, and model-free mammography tools for retrieving task knowledge, inspecting regions of interest, comparing paired and contralateral views, and gathering structured evidence. 

\item \textbf{Skill Evolution:} We adapt an offline skill-evolution workflow that distills reusable reasoning guidance from failed trajectories and retrieves these skills in subsequent cases, allowing us to study experience-driven adaptation without updating the backbone model weights.

\item \textbf{Evaluation:} We empirically investigate \texttt{MammoClaw} on BI-RADS assessment and breast density estimation. Our experiments show that tool augmentation alone does not necessarily improve performance, while skill evolution can improve performance and tool-use behavior.
\end{enumerate}

\begin{figure}[t]
\centering
\includegraphics[width=\linewidth]{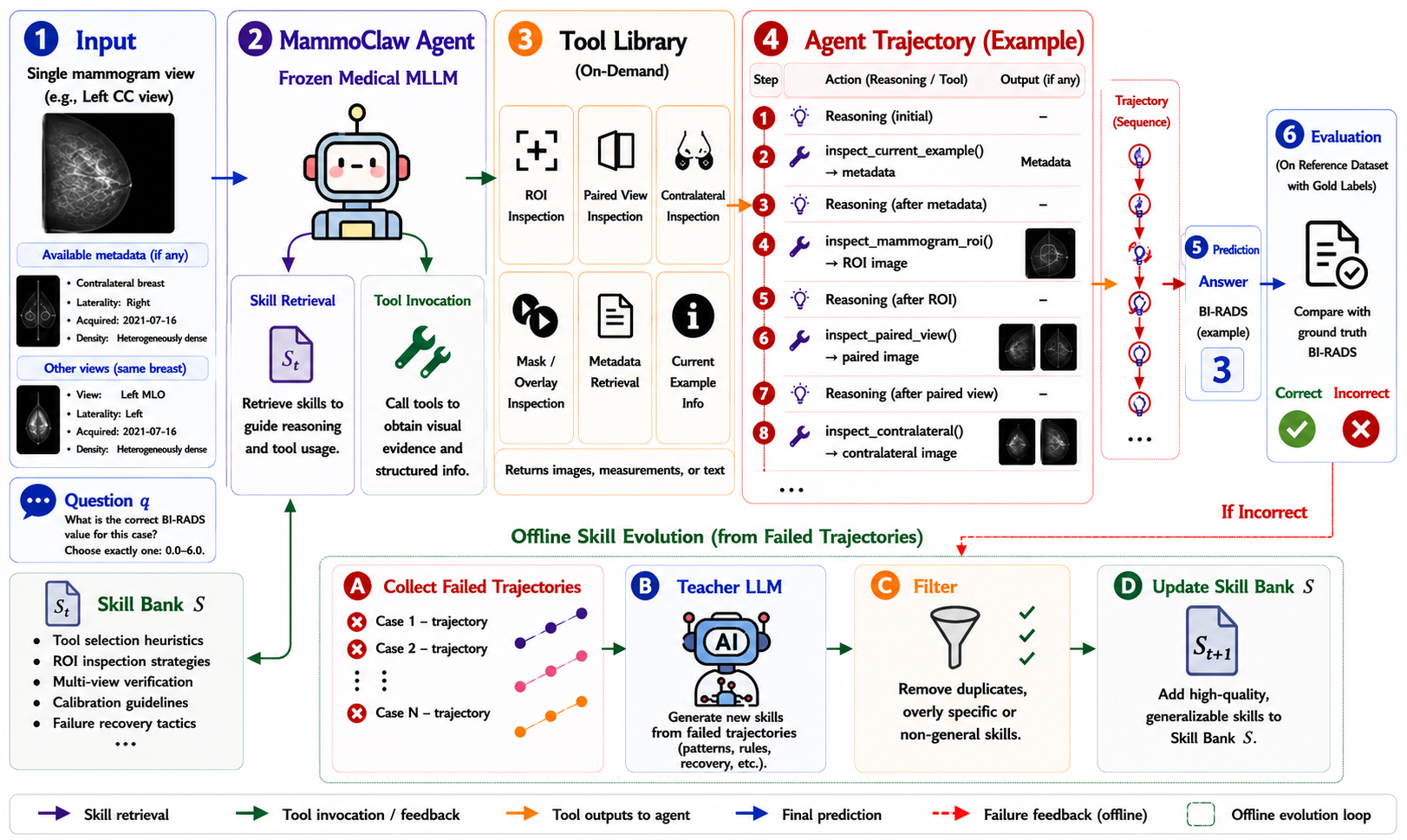}
\caption{Overview of the MammoClaw framework. The top panel illustrates the agent orchestration, while the bottom panel shows the skill-evolution workflow.}
\label{fig:framework}
\end{figure}

\section{MammoClaw Framework}

Figure~\ref{fig:framework} illustrates \texttt{MammoClaw}, a tool-augmented agent framework for exploring breast mammography tasks. Building on an established agentic paradigm, \texttt{MammoClaw} combines a frozen MLLM for multi-step reasoning (Sec.~\ref{sec:agent_orchestration}) with deterministic mammography tools for structured evidence acquisition (Sec.~\ref{sec:mammogram_tools}). To investigate whether prior experience can improve reasoning and tool use without updating model weights, we further incorporate an offline skill-evolution workflow (Sec.~\ref{sec:skill_evolution}) that analyzes failed trajectories from a reference set and distills them into reusable textual skills for subsequent agent execution. We describe each component below.

\subsection{Agent Orchestration}
\label{sec:agent_orchestration}

Following the ReAct framework~\cite{yao2022react}, the orchestration layer alternates between reasoning, tool invocation, and observation before producing a final prediction. Let a mammography case be represented as
$x=(I,q,\mathcal{C})$,
where $I$ denotes the input mammographic image, $q$ denotes the task-specific question, and $\mathcal{C}$ denotes the candidate answer set. Let $f_{\theta}$ denote the frozen MLLM, $\mathcal{A}$ denote the mammography tool library, and $\mathcal{S}$ denote the skill bank. Before the first reasoning step, a lightweight retriever $R(\cdot)$ selects relevant skills
\[
S_x = R(q,\mathcal{S}),
\]
which are prepended to the initial prompt context $\mathcal{H}_0$ together with the input case (see Appendix~\ref{sec:agent_prompt}). At reasoning step $t$, the MLLM receives the current prompt context $\mathcal{H}_{t-1}$ and either produces a final prediction $\hat{y}\in\mathcal{C}$ or selects a tool call
\[
u_t=(A_t,p_t),
\]
where $A_t\in\mathcal{A}$ denotes the selected tool and $p_t$ contains tool-specific inputs (e.g., ROI coordinates). Executing the selected tool returns an observation
\[
o_t=A_t(x,p_t).
\]
The observation is then appended to the prompt context according to
\[
\mathcal{H}_t
=
\mathcal{H}_{t-1}
\oplus
(u_t,o_t),
\]
where $\oplus$ denotes appending the tool invocation and its resulting observation to the prompt context. \texttt{MammoClaw} records the resulting reasoning trajectory as
\[
\tau
=
\left\{
\left(\mathcal{H}_{t-1},u_t,o_t\right)
\right\}_{t=1}^{k},
\]
providing an explicit record of the agent's reasoning and tool interactions that is subsequently used for offline skill evolution.

\subsection{Mammography Tools}
\label{sec:mammogram_tools}

To provide the agent with mammography-specific evidence, \texttt{MammoClaw} employs a lightweight suite of deterministic, \emph{model-free} tools (see Appendix~\ref{sec:toolssuite}), avoiding auxiliary learned modules such as segmentation or detection networks. These include task-context tools, single-image ROI inspection, cross-view and contralateral comparison, and lightweight image-processing utilities. Depending on the operation, they return either textual outputs (e.g., BI-RADS definitions, domain knowledge, metadata) or visual outputs (e.g., ROI crops, paired-view composites, contralateral comparisons), allowing the agent to iteratively gather complementary evidence before producing a final prediction. However, access to these tools alone does not guarantee effective use: the agent must still determine when to invoke them, which tools to select, and how to interpret their outputs. This motivates our investigation of whether prior failed trajectories can provide reusable guidance for subsequent tool use and reasoning.

\subsection{Skill Evolution}
\label{sec:skill_evolution}

Rather than updating model weights, we explore whether failures can be converted into reusable guidance through an offline skill-evolution workflow based on recent agent-learning approaches~\cite{xia2026skillrl,yang2026autoskill}. We apply this workflow to mammography by distilling reusable skills from failed reasoning trajectories.
The evolved skills are stored in a skill bank and retrieved during subsequent inference, providing a lightweight mechanism for experience-driven behavioral adaptation while leaving the underlying MLLM unchanged.

Importantly, these skills are not trained parameters or task-specific model updates; they are reusable textual policies mined from prior failures and injected into the agent context at inference time. Skill evolution proceeds on a labeled reference dataset held out from test evaluation,
\[
\mathcal{D}_{\mathrm{ref}}
=
\left\{
(x_j,y_j)
\right\}_{j=1}^{N}.
\]
Let $\mathcal{S}^{(0)}$ denote the initial skill bank, which is empty in our case. At evolution round $r$, \texttt{MammoClaw} evaluates the reference set using $\mathcal{S}^{(r)}$, producing predictions $\hat{y}_j^{(r)}$ and reasoning trajectories $\tau_j^{(r)}$. Failed trajectories are collected into
\[
\mathcal{T}_{\mathrm{fail}}^{(r)}
=
\left\{
(\tau_j^{(r)},x_j,y_j)
\;\middle|\;
\hat{y}_j^{(r)}\neq y_j
\right\}.
\]

A teacher model $G$ analyzes \emph{all} the failed trajectories together to generate candidate skills,
\[
\widetilde{\mathcal{S}}^{(r)}
=
G\!\left(
\mathcal{T}_{\mathrm{fail}}^{(r)}
\right),
\]
using a dedicated skill-evolution prompt (see Appendix~\ref{sec:skillevolution_prompt}). The generated skills (see Appendix~\ref{sec:skills}) capture reusable reasoning strategies, tool-use patterns, and corrections for recurring failure modes. After semantic-similarity filtering removes redundant candidates, the remaining novel skills $\Delta\mathcal{S}^{(r)}$ are incorporated into the skill bank:
\[
\mathcal{S}^{(r+1)}
=
\mathcal{S}^{(r)}
\cup
\Delta\mathcal{S}^{(r)}.
\]

\section{Experiments}

\paragraph{\bf Implementation.}
We evaluate our framework on Mammo-Bench~\cite{bhole2025mammo} for two mammography assessment tasks: BI-RADS assessment and breast density assessment. We use Qwen3.5-35B-A3B~\cite{qwen3.5} as the underlying agent backbone and report macro F1-score together with 95\% bootstrap confidence intervals. Following the protocol of~\cite{bhole2025mammo}, we randomly split each source dataset into 80\% training and 20\% evaluation sets. This results in an evaluation set of 448 exams from KAU-BCMD~\cite{alsolami2021king} for BI-RADS assessment and 108 exams from DMID~\cite{oza2024digital} for breast density assessment. For skill evolution, we randomly sample $100$ examples from the training set together with their ground-truth labels as the reference set. Skill evolution is performed using DeepSeek-V4-Flash~\cite{xu2026deepseek} as the teacher LLM, with the maximum number of skill-evolution iterations set to $1$ and at most $10$ new skills generated per iteration. Further details of the experimental setup are provided in Appendix~\ref{sec:setup}.

\subsection{Main Results}

\paragraph{\bf Tools Alone Do Not Consistently Improve Performance.}
Table~\ref{tab:ablation_tools_skills} compares the baseline agent, the agent augmented with mammography tools, and the tool-augmented agent with evolved skills. Adding tools results in only small changes in macro F1, from $0.108$ to $0.106$ for BI-RADS assessment.
However, the change is not statistically significant (Appendix~\ref{sec:statistical_analysis}), suggesting that access to mammography-specific tools alone does not consistently improve performance.

\begin{table}[t]
\centering

\begin{subtable}[t]{0.6\linewidth}
\centering
\small
\setlength{\tabcolsep}{3pt}
\renewcommand{\arraystretch}{1.1}
\resizebox{\linewidth}{!}{%
\begin{tabular}{lccc}
\toprule
\textbf{Method} & \textbf{Tools} & \textbf{Skills} & \textbf{Macro-F1} {\textbf{[95\% CI]}} \\
\midrule
Baseline           & $\times$     & $\times$     & {0.108 [0.103, 0.113]} \\
\texttt{MammoClaw} & $\checkmark$ & $\times$     & {0.106 [0.093, 0.122]} \\
\texttt{MammoClaw} & $\checkmark$ & $\checkmark$ & {0.148 [0.121, 0.180]} \\
\bottomrule
\end{tabular}
}
\caption{BI-RADS (KAU-BCMD)}
\end{subtable}

\caption{Effect of tools and evolved skills on mammography assessment. Tools alone do not consistently improve macro-F1, whereas skill evolution leads to improved performance. %
}
\label{tab:ablation_tools_skills}
\end{table}

\paragraph{\bf Skill Evolution Shows Promising Improvements.}
Adding evolved skills increases BI-RADS macro F1 from $0.106$ to $0.148$
relative to the tools-only setting. The improvements are statistically significant (paired bootstrap, Holm-corrected $p<0.05$, see Appendix~\ref{sec:statistical_analysis}). These results suggest that the benefit comes not from tool access alone, but from the additional guidance provided by the evolved skills. However, these results do not establish that the generated skills themselves encode clinically validated reasoning, and further evaluation is needed to assess their robustness and transferability.

\begin{figure}[!ht]
\centering
\begin{minipage}[c]{0.62\linewidth}
\centering
\includegraphics[width=\linewidth]{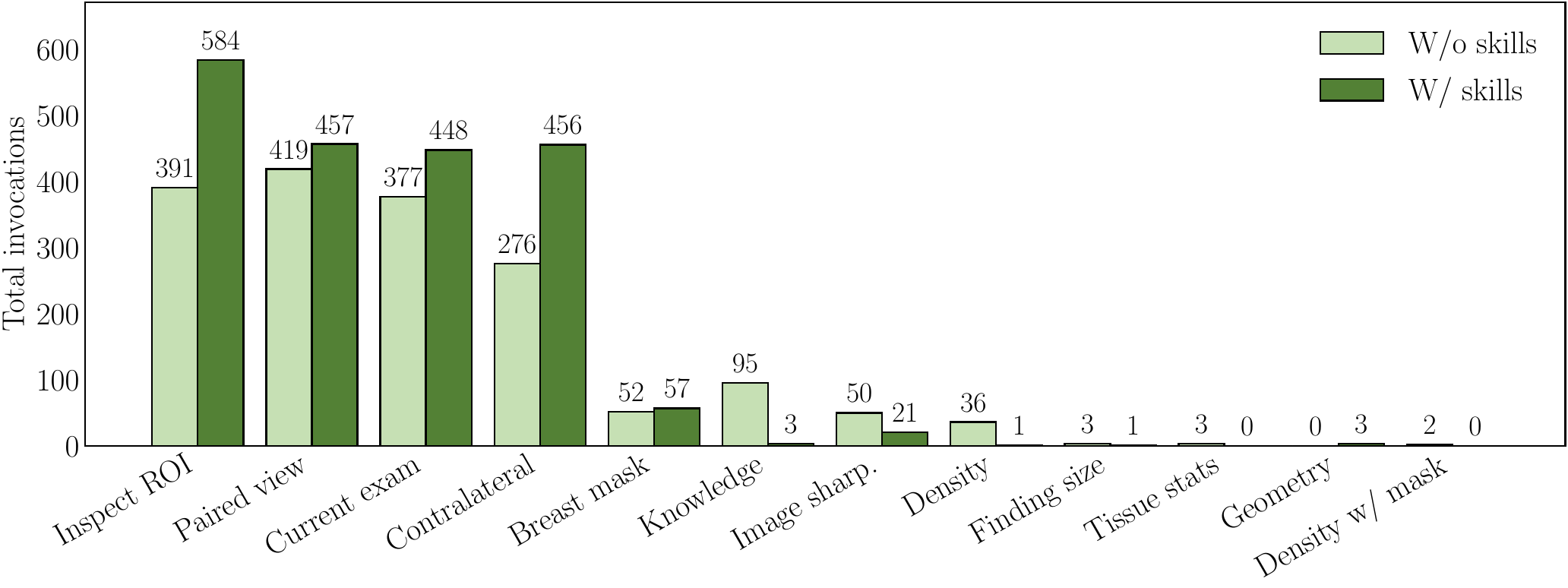}
\end{minipage}
\hfill
\begin{minipage}[c]{0.34\linewidth}
\captionof{figure}{Skill evolution increases evidence gathering per case. After skill evolution, the agent makes more tool calls on the BI-RADS assessment task, suggesting a more deliberate multi-step inspection process before final prediction.}
\label{fig:tool_calls}
\end{minipage}
\end{figure}

\begin{figure}[!ht]
\centering
\begin{minipage}[c]{0.62\linewidth}
\centering
\includegraphics[width=\linewidth]{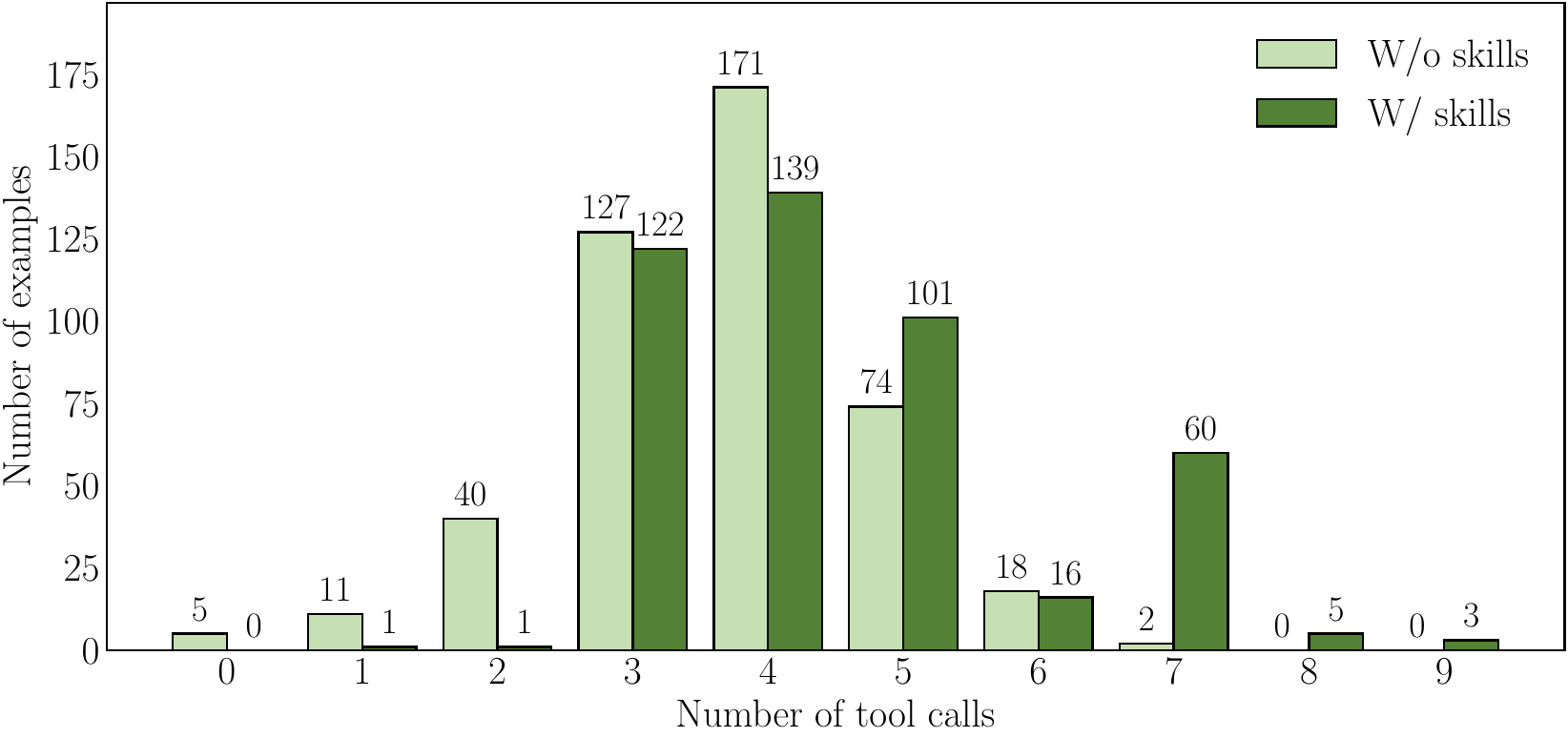}
\end{minipage}
\hfill
\begin{minipage}[c]{0.34\linewidth}
\captionof{figure}{Comparison of tool-call frequency per case before and after skill evolution on the BI-RADS assessment task.}
\label{fig:tool_freq_birads}
\end{minipage}
\end{figure}

\paragraph{\bf Insights on Evolved Skills and Tool Usage.}
Figure~\ref{fig:tool_calls} illustrates changes in the agent's interaction with the mammography tool suite following skill evolution. The generated skills (Appendix~\ref{sec:skills}) capture reusable reasoning strategies rather than explicit tool-selection rules, including systematic inspection of suspicious regions, cross-view verification, sequential resolution of uncertainty, and grounding conclusions in tool observations. Following skill evolution, the agent invokes ROI inspection and contralateral breast comparison tools more frequently. These changes are consistent with the reasoning strategies reflected in the evolved skills. 
Moreover, we observe fewer calls to the knowledge-retrieval tool after skill evolution, suggesting that the evolved skills may provide some of the task-relevant guidance that the agent would otherwise seek through knowledge retrieval. Figure~\ref{fig:tool_freq_birads} further shows that evolved skills generally lead to more tool calls per case across both BI-RADS and breast density assessment, suggesting more extensive evidence gathering during inference. These observations provide an initial indication that skill evolution can alter tool-use behavior, while the relationship between tool usage and prediction quality remains an open question.

\begin{figure}[!ht]
        \centering
\includegraphics[width=0.85\linewidth,
    ]{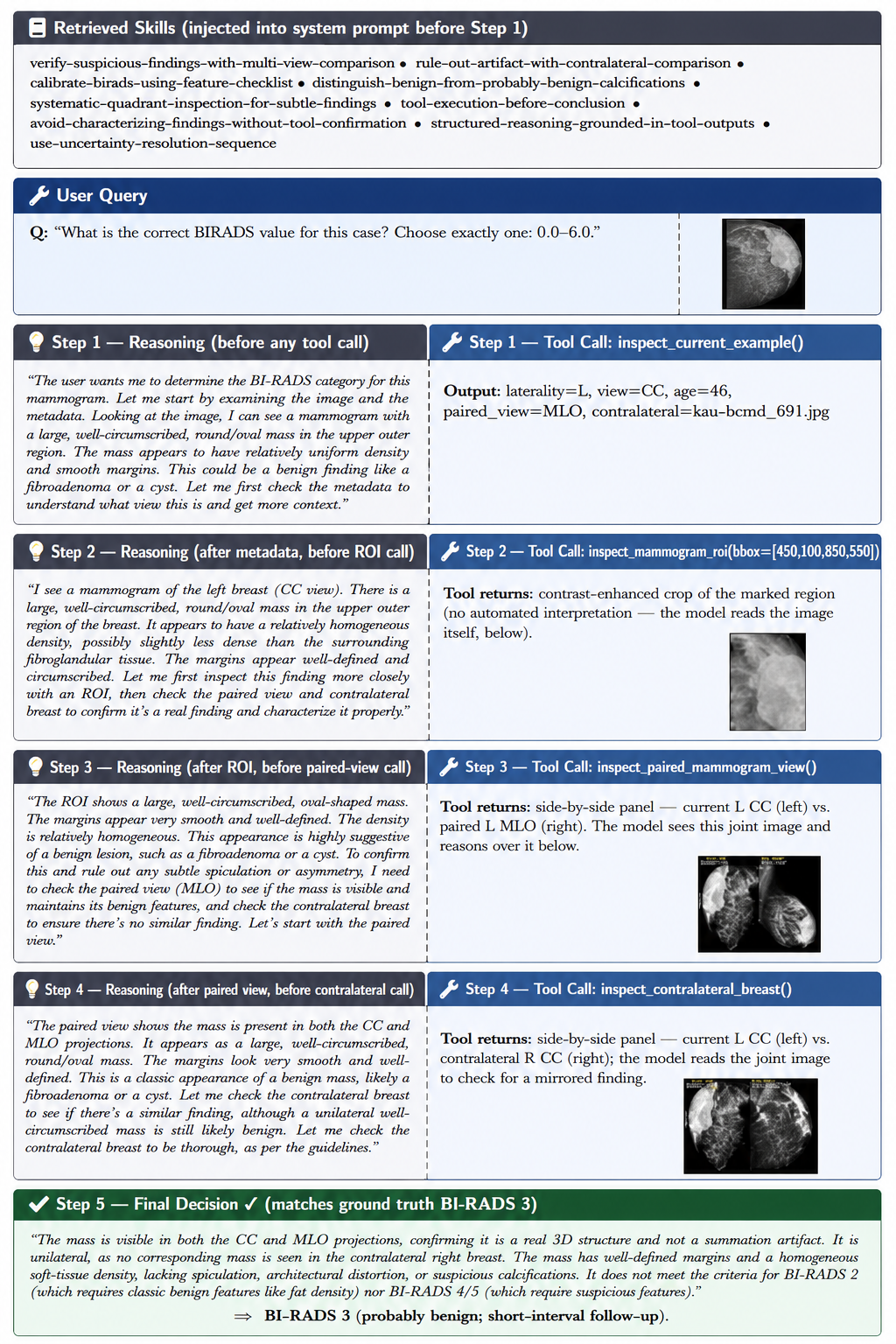}
\caption{\textbf{Skill-guided agent trajectory for BI-RADS prediction.} The agent first localizes the suspicious region, verifies the finding using the paired MLO view and the contralateral breast, and incrementally refines its reasoning before producing the final assessment. The trajectory illustrates how tool interactions and evolved skills can support evidence-driven mammography assessment.}
\label{fig:trajectory}
\end{figure}

\paragraph{\bf Trajectory Visualization.}
Figure~\ref{fig:trajectory} visualizes a representative reasoning trajectory generated by \texttt{MammoClaw} from a CC-view mammogram. The agent first localizes a suspicious region, verifies the finding using the paired MLO view and the contralateral breast, and incrementally refines its reasoning before producing the final assessment. This example illustrates how the framework exposes tool interactions and intermediate evidence during mammography assessment and provides a qualitative example of how evolved skills may guide evidence acquisition.

\section{Conclusion}

In this paper, we investigated \texttt{MammoClaw}, a training-free, tool-augmented agent framework for breast mammography. \texttt{MammoClaw} combines a frozen MLLM with deterministic mammography-specific tools and an offline skill-evolution workflow that distills reusable textual guidance from failed trajectories without updating model weights. Across two mammography assessment tasks, we find that tool augmentation alone does not consistently improve performance in our setting, whereas adding failure-derived skills is associated with promising gains relative to the tools-only setting. The accompanying changes in tool-use behavior further suggest that skill evolution can influence how the agent gathers and integrates evidence during inference. Overall, these results provide an initial indication that failure-driven skill evolution is a promising direction for developing domain-specialized agent harnesses for mammography. We hope \texttt{MammoClaw} serves as a useful testbed for further investigation of mammography-specific tools, skill evolution, and agent behavior. We refer the reader to the Supplementary Material for limitations of our work and further details on the experimental settings, prompts, and evolved skills.

\noindent \textbf{Disclosure of Interests.}
The author declares no competing interests relevant to the content of this article.

\bibliographystyle{splncs04}
\bibliography{references}

\appendix

\section{Limitations and Future Work}
\label{sec:limitations}

\texttt{MammoClaw} is an initial exploratory study of skill-evolving agent frameworks for mammography. Our evaluation is limited to two assessment tasks, BI-RADS prediction and breast density estimation, using a single frozen MLLM backbone and one benchmark dataset per task. In addition, Qwen3.5-35B-A3B shows modest performance in the no-tools setting, which may limit the extent to which the benefits of the agent harness can be assessed. Future evaluations with stronger mammography-capable backbones, additional datasets, and broader tasks, such as abnormality classification, are needed to better characterize the generality and robustness of these findings.

The evolved skills are derived from a labeled reference set and are not independently validated by clinical experts. While evolved skills are associated with performance improvements in our experiments, their clinical relevance, robustness, and transferability across datasets remain open questions. Future work should therefore evaluate skill quality more directly, including whether the generated skills align with expert mammography reasoning, whether they encode clinically meaningful guidance, and whether they remain useful under dataset shift.

Another important direction is a systematic analysis of the agent harness itself. In particular, the effects of the system prompt, tool descriptions, number of injected failure trajectories, skill-generation prompt, teacher LLM, skill-validation strategy, and skill-retrieval mechanism remain unexplored. Controlled ablations of these components, together with step-level and trajectory-level metrics, could provide a more detailed understanding of how skill evolution changes evidence acquisition, tool use, and final prediction behavior. Such analyses may also help distinguish improvements arising from the skills themselves from those arising from changes in prompting or tool interaction patterns.

Finally, our current skill-evolution setup uses a single evolution iteration and a relatively small reference set. This provides a controlled starting point for studying experience-driven adaptation, but does not establish how skill evolution behaves over longer horizons or at larger scales. Future work could investigate multi-round evolution, skill pruning and validation, skill transfer across tasks and datasets, and mechanisms for preventing the accumulation of redundant or potentially misleading guidance.

\section{Experimental Setup}\label{sec:setup}
\subsection{Dataset Details}\label{sec:dataset}

 We evaluate our method on Mammo-Bench~\cite{bhole2025mammo} for two mammography assessment tasks: BI-RADS assessment and breast density assessment. Following the protocol of~\cite{bhole2025mammo}, we randomly split each dataset into 80\% training and 20\% evaluation sets. From the training split, we randomly select 100 examples together with their ground-truth labels as the reference set for skill evolution. In particular, we use KAU-BCMD~\cite{alsolami2021king} for BI-RADS assessment and DMID~\cite{oza2024digital} for breast density assessment. KAU-BCMD provides paired CC and MLO views together with contralateral breast images, enabling evaluation of all proposed tools. In contrast, DMID contains only single-view mammograms.

For BI-RADS assessment, the evaluation set contains 448 examples distributed across BI-RADS categories 1, 3, 4, and 5, with 367, 59, 18, and 4 examples, respectively; categories 0, 2, and 6 are absent. The corresponding 100-example reference set contains 83, 14, and 3 examples from BI-RADS categories 1, 3, and 5, respectively. For breast density assessment, the evaluation set contains 108 examples distributed across density categories A, B, C, and D, with 18, 39, 42, and 9 examples, respectively. The corresponding 100-example reference set contains 18, 39, 38, and 5 examples from density categories A, B, C, and D, respectively.

\subsection{Additional Implementation Details}\label{sec:implementation_details}

We used Qwen3.5-35B-A3B~\cite{qwen3.5} as the underlying agent backbone and DeepSeek-V4-Flash~\cite{xu2026deepseek} as the teacher LLM for skill evolution. The maximum number of skill-evolution iterations was set to 1, with at most 10 new skills generated per iteration.  We used the OpenRouter interface to access Qwen3.5-35B-A3B and DeepSeek-V4-Flash. 
During skill generation, we passed a maximum of $50$ failed trajectories at a time. After the teacher LLM proposes new skills from failed trajectories, MammoClaw filters near-duplicates using both name-token Jaccard similarity and semantic cosine similarity over full-skill embeddings; accepted skills are then added to the dynamic skill bank. At inference time, we retrieve all generated task-relevant skills from the bank without additional filtering, so the full relevant skill set is injected into the agent context.

\section{Additional Results}\label{sec:statistical_analysis}
In this section, we present additional results for breast density estimation and report the corresponding statistical analyses.
We evaluate statistical significance using two complementary tests (Table~\ref{tab:stats}). First, we compare paired predictions using McNemar's test. Second, we compare macro-F1 using paired bootstrap resampling (5{,}000 resamples), reporting both 95\% confidence intervals and paired-bootstrap $p$-values. We apply Holm--Bonferroni correction across the three pairwise comparisons for each task.

Overall, both tasks show a similar trend: skill evolution improves performance, whereas adding tools alone does not lead to statistically significant improvements in macro-F1. For BI-RADS, skill evolution significantly improves macro-F1 compared with both the no-tool and tools-only settings (bootstrap $p<0.001$ for both comparisons). McNemar's test likewise indicates significant differences for both comparisons ($p<0.001$). In contrast, adding tools without skills does not significantly change macro-F1 (bootstrap $p=0.70$), although McNemar's test indicates a significant difference ($p=0.0017$).

For breast density ($n=108$), skill evolution significantly improves macro-F1 over the tools-only setting (bootstrap $p=0.017$), whereas the comparison with the no-tool baseline is not statistically significant (bootstrap $p=0.18$). McNemar's test is significant only for the no-tool versus tools+skills comparison ($p<0.001$); the remaining comparisons are not statistically significant ($p=0.062$ for both).

\begin{table}[h]
\centering
\small
\resizebox{\linewidth}{!}{%
\setlength{\tabcolsep}{10pt}
\begin{tabular}{llccc}
\toprule
\textbf{Task} & \textbf{Comparison} & $\Delta$\textbf{Macro-F1} & \textbf{Boot. $p$ (Holm)} & \textbf{McNemar (acc.) $p$ (Holm)} \\
\midrule
BI-RADS & No tools $\rightarrow$ Tools  & $-0.002$ & $0.70$   & $0.0017$       \\
BI-RADS & No tools $\rightarrow$ Skills & $+0.040$ & $<0.001$       & $<0.001$       \\
BI-RADS & Tools $\rightarrow$ Skills    & $+0.043$ & $<0.001$       & $<0.001$       \\
\midrule
Density & No tools $\rightarrow$ Tools  & $+0.019$ & $0.70$  & $0.062$  \\
Density & No tools $\rightarrow$ Skills & $+0.093$ & $0.18$   & $<0.001$       \\
Density & Tools $\rightarrow$ Skills    & $+0.074$ & $0.017$        & $0.062$  \\
\bottomrule
\end{tabular}
}
\vspace{4pt}

\caption{Statistical significance of macro-F1 differences across tool and skill configurations.}
\label{tab:stats}
\end{table}

\section{Tool Suite}\label{sec:toolssuite}

We show one representative call per tool below, drawn directly from logged agent trajectories.

\begin{center}

\begin{toolmeta}{Tool: \texttt{inspect\_current\_example}}
\scriptsize
\textbf{Category:} Task context\\[2pt]
\textbf{Description:} Return the current example metadata.\\[2pt]
\textbf{Input:} \texttt{(no arguments)}\\[4pt]
\textbf{Output:} \texttt{source\_dataset=kau-bcmd, laterality=R, view=MLO, subject\_age=61.0, has\_paired\_view=true, paired\_view=CC, contralateral\_image\_path=\ldots} 
\end{toolmeta}

\begin{toolmeta}{Tool: \texttt{retrieve\_knowledge}}
\scriptsize
\textbf{Category:} Task context\\[2pt]
\textbf{Description:} Retrieve task-specific domain knowledge for the current task, such as BI-RADS category definitions for BI-RADS assessment or density-band definitions for density assessment.\\[2pt]
\textbf{Input:} \texttt{(no arguments)}\\[4pt]
\textbf{Output:} task-specific label guidance --- for the BI-RADS task, the full category definitions are returned verbatim:
\begin{itemize}\itemsep0pt\parskip0pt\topsep2pt
\item \textbf{BI-RADS 0 (Incomplete):} Additional imaging evaluation and/or prior examinations are needed before a final assessment can be made.
\item \textbf{BI-RADS 1 (Negative):} No suspicious findings.
\item \textbf{BI-RADS 2 (Benign):} Benign finding with no evidence of malignancy.
\item \textbf{BI-RADS 3 (Probably Benign):} Very low likelihood of malignancy ($<$2\%); short-interval follow-up is typically recommended.
\item \textbf{BI-RADS 4 (Suspicious):} Suspicious abnormality; tissue diagnosis (biopsy) should be considered.
  \begin{itemize}\itemsep0pt\parskip0pt\topsep2pt
  \item \textbf{4A:} Low suspicion for malignancy.
  \item \textbf{4B:} Moderate suspicion for malignancy.
  \item \textbf{4C:} High suspicion for malignancy, but not classic for cancer.
  \end{itemize}
\item \textbf{BI-RADS 5 (Highly Suggestive of Malignancy):} Very high likelihood of malignancy ($>$95\%); appropriate action should be taken.
\item \textbf{BI-RADS 6 (Known Biopsy-Proven Malignancy):} Malignancy confirmed by prior biopsy.
\end{itemize}
\end{toolmeta}

\begin{toolstep}{Tool: \texttt{inspect\_mammogram\_roi}}{2.4cm}
\scriptsize
\textbf{Category:} Single-image inspection\\[2pt]
\textbf{Description:} Crop and enlarge a suspicious region from the current mammogram using normalized 0--1000 image-grid coordinates.\\[2pt]
\textbf{Input:} \texttt{bbox=[200, 200, 800, 800]}, \texttt{description=\textquotedblleft central breast region to assess whether dense tissue appears as scattered islands (B) or confluent/heterogeneous regions (C)\textquotedblright}, \texttt{enhance\_contrast=true}\\[4pt]
\textbf{Output:} ROI crop attached (right)
\tcblower
\includegraphics[height=2.2cm]{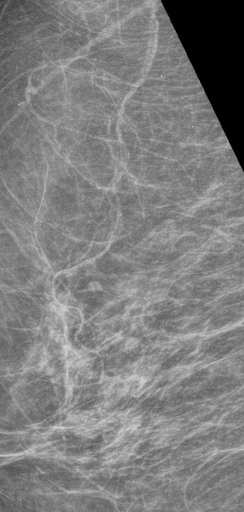}
\end{toolstep}

\begin{toolstep}{Tool: \texttt{measure\_finding\_size}}{2.4cm}
\scriptsize
\textbf{Category:} Single-image measurement\\[2pt]
\textbf{Description:} Measure the width and height of a suspicious finding by drawing annotated measurement bars on a contextual crop.\\[2pt]
\textbf{Input:} \texttt{bbox=[0, 200, 250, 550]}, \texttt{description=\textquotedblleft spiculated mass at nipple/areolar region\textquotedblright}\\[4pt]
\textbf{Output:} annotated crop with measurement bars (right).
\tcblower
\includegraphics[height=2.2cm]{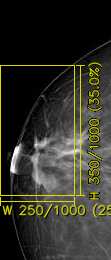}
\end{toolstep}

\begin{toolmeta}{Tool: \texttt{estimate\_breast\_density}}
\scriptsize
\textbf{Category:} Single-image density\\[2pt]
\textbf{Description:} Estimate fibroglandular density of the current mammogram using dual Otsu thresholding over breast and dense-tissue regions.\\[2pt]
\textbf{Input:} \texttt{(no arguments)}\\[4pt]
\textbf{Output:} \texttt{density\_percent=35.5}, dual-Otsu thresholding (background$\rightarrow$breast, then breast$\rightarrow$dense tissue); returned with an explicit caveat not to assign an A/B/C/D category from the percentage alone.
\end{toolmeta}

\begin{toolmeta}{Tool: \texttt{measure\_image\_sharpness}}
\scriptsize
\textbf{Category:} Single-image quality\\[2pt]
\textbf{Description:} Score technical sharpness via Laplacian variance and Tenengrad gradient energy, returning a sharpness rating.\\[2pt]
\textbf{Input:} \texttt{(no arguments)}\\[4pt]
\textbf{Output:} \texttt{laplacian\_variance=134.64, tenengrad=1287.97, sharpness\_rating=\textquotedblleft acceptable\textquotedblright} on a $2146\times2977$ image.
\end{toolmeta}

\begin{toolmeta}{Tool: \texttt{compute\_tissue\_statistics}}
\scriptsize
\textbf{Category:} Single-image texture\\[2pt]
\textbf{Description:} Compute intensity histogram statistics over the breast foreground while excluding background via Otsu segmentation.\\[2pt]
\textbf{Input:} \texttt{(no arguments)}\\[4pt]
\textbf{Output:} \texttt{mean\_intensity=136.83, std\_intensity=27.62, entropy\_bits=6.767, skewness=$-$0.431} over the breast foreground (background excluded via Otsu).
\end{toolmeta}

\begin{toolstep}{Tool: \texttt{inspect\_paired\_mammogram\_view}}{2.4cm}
\scriptsize
\textbf{Category:} Cross-view comparison\\[2pt]
\textbf{Description:} Inspect the alternate projection of the same breast by returning paired CC and MLO views side by side.\\[2pt]
\textbf{Input:} \texttt{(no arguments)}\\[4pt]
\textbf{Output:} side-by-side panel, current L~CC (left) vs.\ paired L~MLO (right)
\tcblower
\includegraphics[height=2.2cm]{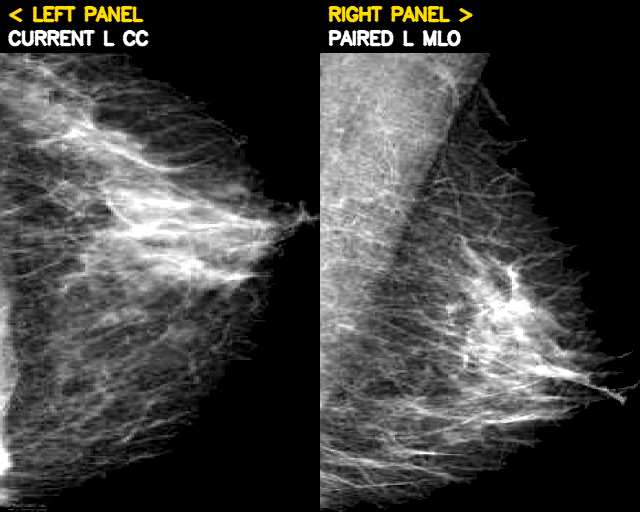}
\end{toolstep}

\begin{toolstep}{Tool: \texttt{inspect\_contralateral\_breast}}{2.4cm}
\scriptsize
\textbf{Category:} Cross-breast comparison\\[2pt]
\textbf{Description:} Compare the current breast with the contralateral breast from the same subject, source dataset, and view projection.\\[2pt]
\textbf{Input:} \texttt{(no arguments)}\\[4pt]
\textbf{Output:} side-by-side panel, current L~MLO (left) vs.\ contralateral R~MLO (right), for global asymmetry comparison.
\tcblower
\includegraphics[height=2.2cm]{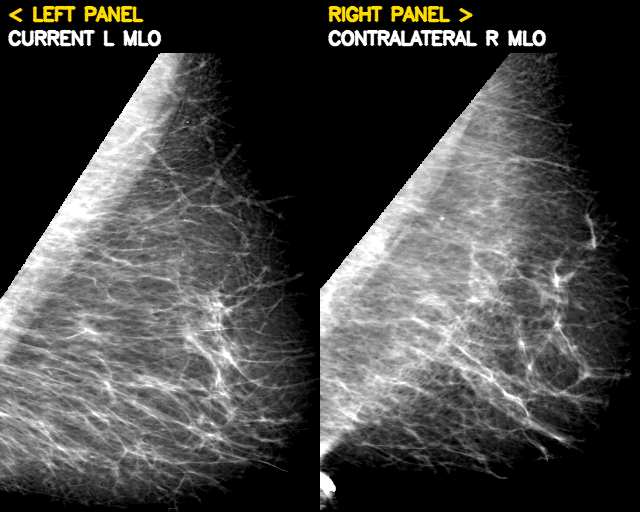}
\end{toolstep}

\end{center}

\section{Qualitative Results}

Figure~\ref{fig:trajectory_example_birads4} shows a representative BI-RADS trajectory in which the tool-augmented agent with evolved skills reaches a correct suspicious assessment. The case illustrates how the agent uses the tool suite to gather complementary evidence: an ROI inspection localizes architectural distortion, paired-view comparison verifies that the finding persists across projections, and contralateral comparison rules out a symmetric normal variant. The final prediction is grounded in these tool observations rather than a single-pass interpretation of the input image.

\begin{center}

\begin{tcolorbox}[
  enhanced, breakable,
  colback=trajthink, colframe=trajthinkframe, boxrule=0.5pt, arc=1pt,
  left=6pt, right=6pt, top=3pt, bottom=3pt,
  before skip=4pt, after skip=4pt,
  title={\faBook~Retrieved Skills},
  coltitle=white, colbacktitle=trajthinkframe,
  fonttitle=\bfseries\footnotesize\sffamily,
  fontupper=\scriptsize\raggedright
]
\texttt{verify-suspicious-findings-with-multi-view-comparison} \textbullet\ 
\texttt{rule-out-artifact-with-contralateral-comparison} \textbullet\ 
\texttt{structured-reasoning-grounded-in-tool-outputs} \textbullet\ 
\texttt{calibrate-birads-using-feature-checklist}
\end{tcolorbox}

\begin{toolstep}{User Query}{2.2cm}
\scriptsize
\textbf{Q:} ``What is the correct \texttt{BI-RADS} value for this case? Choose exactly one: 0.0--6.0.''
\tcblower
\includegraphics[height=2.0cm]{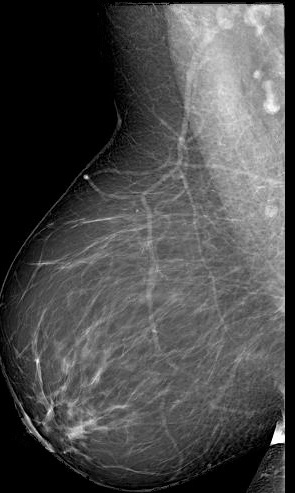}
\end{toolstep}

\begin{thinkstep}{Step 1 --- Initial Reasoning}
\scriptsize\textit{``The user wants me to determine the BI-RADS category for this mammogram. Let me start by examining the image and then use the appropriate tools to make a thorough assessment. I should first inspect the current example to understand the metadata, then systematically evaluate the breast.''}
\end{thinkstep}

\begin{toolmeta}{Step 1 --- Tool Call: \texttt{inspect\_current\_example()}}
\scriptsize
\textbf{Output:} \texttt{source\_dataset=kau-bcmd, laterality=R, view=MLO, subject\_age=45.0, has\_paired\_view=true, paired\_view=CC, contralateral\_image\_path=kau-bcmd\_1004.jpg}
\end{toolmeta}

\begin{thinkstep}{Step 2 --- Reasoning (before ROI call)}
\scriptsize\textit{``I see an MLO view of the right breast. There appears to be an area of architectural distortion or asymmetry in the lower portion of the breast. Let me first inspect this area more closely with an ROI, then check the paired view and contralateral breast.''}
\end{thinkstep}

\begin{toolstep}{Step 2 --- Tool Call: \texttt{inspect\_mammogram\_roi(bbox=[200,650,450,950])}}{2.7cm}
\scriptsize
\textbf{Input:} \texttt{description=\textquotedblleft area of architectural distortion/asymmetry in lower breast\textquotedblright}\\[2pt]
\textbf{Tool returns:} contrast-enhanced crop of the lower-breast region; the model identifies radiating lines converging toward a central point.
\tcblower
\includegraphics[height=2.4cm]{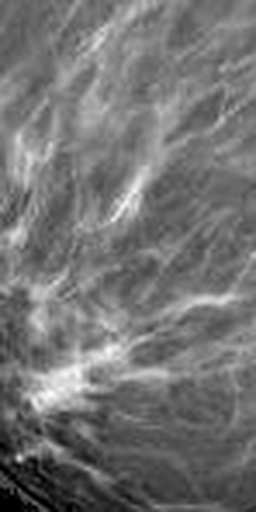}
\end{toolstep}

\begin{thinkstep}{Step 3 --- Reasoning (after ROI, before paired-view call)}
\scriptsize\textit{``The ROI shows an area of architectural distortion with spiculations radiating from a central point near the skin/chest wall area. This looks suspicious. I need to check the paired view to confirm if this is a real finding or an artifact.''}
\end{thinkstep}

\begin{toolstep}{Step 3 --- Tool Call: \texttt{inspect\_paired\_mammogram\_view()}}{3.3cm}
\scriptsize
\textbf{Tool returns:} side-by-side panel, current R~MLO (left) vs.\ paired R~CC (right). The finding is present on both projections, supporting a real 3D finding rather than summation artifact.
\tcblower
\includegraphics[height=2.4cm]{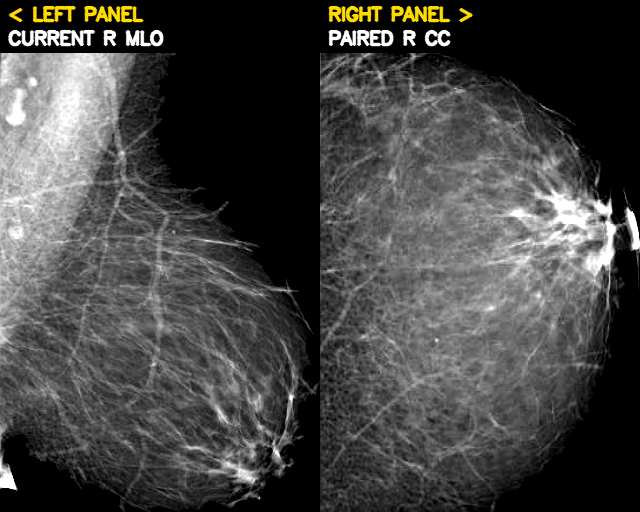}
\end{toolstep}

\begin{thinkstep}{Step 4 --- Reasoning (after paired view, before contralateral call)}
\scriptsize\textit{``The paired view shows the architectural distortion/linear density in the lower posterior aspect of the breast on both the MLO and CC views. This confirms it is a real finding. Now I need to check the contralateral breast to see whether it is a normal bilateral variant.''}
\end{thinkstep}

\begin{toolstep}{Step 4 --- Tool Call: \texttt{inspect\_contralateral\_breast()}}{3.3cm}
\scriptsize
\textbf{Tool returns:} side-by-side panel, current R~MLO (left) vs.\ contralateral L~MLO (right). The opposite breast lacks a corresponding focal distortion, supporting a unilateral suspicious finding.
\tcblower
\includegraphics[height=2.4cm]{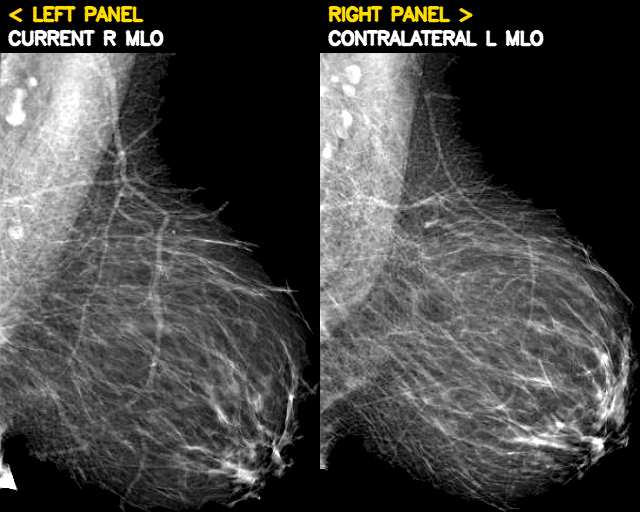}
\end{toolstep}

\begin{finalstep}{Step 5 --- Final Decision \ding{51} (matches ground truth BI-RADS 4)}
\scriptsize\textit{``The mammogram reveals a unilateral area of architectural distortion visible on both MLO and CC views, confirmed by multi-view comparison. The contralateral breast lacks a corresponding finding, indicating it is not a normal variant. Architectural distortion is a suspicious finding that typically warrants a BI-RADS 4 assessment for further evaluation.''} $\Rightarrow$ \textbf{BI-RADS 4}.
\end{finalstep}

\end{center}

\captionof{figure}{\textbf{Skill-guided agent trajectory for a suspicious BI-RADS case.} The agent uses \texttt{inspect\_mammogram\_roi} to localize architectural distortion, \texttt{inspect\_paired\_mammogram\_view} to confirm persistence across projections, and \texttt{inspect\_contralateral\_breast} to rule out a normal bilateral variant before predicting BI-RADS~4.}
\label{fig:trajectory_example_birads4}

\section{Prompts}\label{sec:prompts}

\subsection{System Prompt of the Agent}\label{sec:agent_prompt}
We present the system prompt for the agent below, containing placeholder for retrieved skills.

\begin{promptbox}{promptpurple}
\begin{flushleft}

You are MammoClaw, a mammography task-solving agent.\\

\vspace{0.75em}
Analyze the mammogram supplied by the user for the requested BIRADS task.\\
Use only visible image evidence, the trusted guidance below, and outputs from available tools.\\
Do not infer labels from file paths, dataset identity, record indices, or hidden metadata.

\vspace{0.75em}
Trusted skill guidance:\\
\texttt{\{\{trusted\_skill\_guidance\}\}}

\vspace{0.75em}
Recent lessons:\\
- none

\vspace{0.95em}
Call \texttt{retrieve\_knowledge} at most once to resolve uncertainty about category definitions; do not restate the retrieved definitions unless they directly support the final decision. Use an available tool only when it materially helps. Call \texttt{inspect\_mammogram\_roi} at most 2 times total.\\
Multi-view tools are available for this case: \texttt{inspect\_paired\_mammogram\_view} (to compare CC and MLO projections of the same breast) and \texttt{inspect\_contralateral\_breast} (to assess bilateral symmetry and rule out normal variants). Consider using these when a finding is ambiguous or when bilateral comparison would help confirm or exclude a diagnosis.

\vspace{0.75em}
Tool usage policy:\\
- Do not mention, invent, request, or call any tool whose name is not in the allowed list.\\
- Prefer the minimum number of tool calls needed to reach a confident decision.\\
- Invoke a tool only if its output is expected to influence the final answer.\\
- Avoid redundant or overlapping tool calls.\\
- Do not call additional tools unless they are expected to provide new, decision-relevant evidence.

\vspace{0.75em}
Reasoning policy:\\
- Treat pre-tool observations as hypotheses, not conclusions. Do not characterize a finding before inspecting it with the appropriate tool.\\
- When a tool is clearly needed, call it immediately. Do not describe or justify a tool-use plan before invoking the tool. Do not call \texttt{inspect\_current\_example} unless the metadata is expected to influence the diagnosis or determine which comparison tool to use.\\
- After calling a tool, reason primarily from its output rather than repeatedly reinterpreting the original image.\\
- Do not repeatedly verify the same conclusion using the same evidence.\\
- Do not enumerate answer choices or category definitions unless needed to distinguish between specific candidate labels.\\
- Every reasoning step must introduce new visual evidence, a new comparison, or new tool evidence.\\
- If the current step would only restate or rename previous observations, stop reasoning and return the final answer.\\
- Do not repeatedly describe the same visual structure using different hypotheses unless new evidence distinguishes them.\\
- Once the available evidence sufficiently distinguishes among the answer choices, commit to the best-supported label.

\vspace{0.75em}
When ready, return strict JSON only:\\
\texttt{\{"label":"exact answer choice",}\\
\texttt{"tool\_findings":"REQUIRED: one bullet per tool call, in order called. Every tool you used must appear here. Format: '- tool\_name: key finding from that tool'. If no tools were used write 'no tools used'. Do NOT skip any tool.",}\\
\texttt{"reasoning\_summary":"2-5 concise sentences explaining how the recorded tool outputs support the selected label. Base the explanation solely on the findings listed in tool\_findings. Do not introduce new visual observations or reasoning absent from tool\_findings. Explicitly reference only the tool names you actually called and explain how each cited tool contributed to the final decision."\}}

\vspace{0.75em}
CRITICAL RULES:\\
- \texttt{label} must exactly match one answer choice supplied by the user.\\
- \texttt{tool\_findings} must contain one bullet for EVERY tool you called, in the order you called them. If you called 4 tools, there must be 4 bullets. Omitting any tool call is an error.\\
- \texttt{reasoning\_summary} must be grounded solely in \texttt{tool\_findings}. Do not introduce evidence absent from \texttt{tool\_findings}.\\
Before returning the final JSON, ensure that: every tool call appears once in \texttt{tool\_findings}; \texttt{reasoning\_summary} is supported entirely by \texttt{tool\_findings}.

\end{flushleft}
\end{promptbox}

\subsection{System Prompt of the Skill Teacher}\label{sec:skillevolution_prompt}
We present the system prompt for the skill teacher below, containing the skill-generation policy, failure-analysis policy, and skill-quality requirements. 

\begin{promptbox}{promptpurple}
\begin{flushleft}

\textbf{Role:}\\
You are a skill engineer for a mammography vision-language agent. Your objective is to discover reusable reasoning skills from failed reasoning trajectories that improve future performance across mammography tasks.

\vspace{0.75em}
\textbf{Inputs:}\\
The host provides:\\
\hspace*{1em}- previously failed reasoning trajectories;\\
\hspace*{1em}- the current task name;\\
\hspace*{1em}- the existing skill library;\\
\hspace*{1em}- the maximum number of new skills to generate.

\vspace{0.75em}
\textbf{Failure Analysis:}\\
Analyze the failed trajectories to identify recurring root-cause failures. Determine whether the errors arise from evidence gathering, evidence interpretation, localization, tool usage, verification, calibration, or task-specific reasoning. Focus on high-impact failure modes that generalize across multiple cases rather than isolated mistakes.

\vspace{0.75em}
\textbf{Skill Generation Policy:}\\
Generate a new skill only if it addresses a reusable failure mode that is not already covered by the existing skill library. Each proposed skill should encode a reusable reasoning procedure, inspection strategy, verification procedure, or tool-selection strategy. Avoid duplicating, rephrasing, or making superficial modifications to existing skills.

\vspace{0.75em}
\textbf{Generalization Constraints:}\\
A skill should remain useful even if the anatomy, diagnosis, image, or answer labels differ. Do not encode case-specific corrections, patient details, dataset shortcuts, diagnostic thresholds, or label mappings. The generated guidance must be supported by the observed failures rather than external domain knowledge alone.

\vspace{0.75em}
\textbf{Skill Quality Requirements:}\\
Prefer a small number of high-quality skills that each address multiple failure cases. A strong skill should specify when it applies, what evidence should be collected, what should be verified, how uncertainty should be reduced, and which common failure modes should be avoided before reaching a final decision.

\vspace{0.75em}
\textbf{Output Contract:}\\
Return a JSON array of newly discovered skills. Each skill contains a unique name, a one-sentence description, a category, and structured markdown consisting of a title, reusable reasoning procedure, and an anti-pattern section. Return an empty array if no broadly reusable skill can be inferred.

\end{flushleft}
\end{promptbox}

\section{BI-RADS Evolved Skills}
\label{sec:skills}

This section presents the mammography skills for the BI-RADS task automatically evolved using DeepSeek-V4-Flash as the teacher model. The skills are distilled from failed reasoning trajectories on the reference set and stored in a reusable skill library. Rather than encoding explicit tool-selection policies, they capture reusable reasoning strategies, including systematic evidence gathering, multi-view verification, uncertainty resolution, BI-RADS calibration, and grounding conclusions in tool observations. During inference, these skills are retrieved to guide the agent’s reasoning and evidence integration, enabling behavioral adaptation without updating the underlying MLLM.

\begin{skillbox}
\begin{flushleft}

\textbf{Metadata}\\
name: verify-suspicious-findings-with-multi-view-comparison\\
description: Before concluding a suspicious finding like spiculation or architectural distortion, always compare with the paired view and contralateral breast to rule out artifacts or normal tissue overlap.\\
metadata:\\
\hspace*{1em}category: general\\
\hspace*{1em}source: dynamic\\
\hspace*{1em}task: BI-RADS

\vspace{0.75em}
\textbf{Verify Suspicious Findings with Multi-view Comparison}\\

1. Immediately invoke \texttt{inspect\_paired\_mammogram\_view} to see if the finding is visible in both CC and MLO projections. If it is not clearly present in both, it may be a summation artifact.\\
2. Then invoke \texttt{inspect\_contralateral\_breast} to check for bilateral symmetry. A finding that appears in the same location on the opposite breast is likely a normal anatomic variant.\\
3. Only after confirming the finding is unilateral and present on two orthogonal views should you assign a BI-RADS category of 4 or 5.\\
4. If the finding disappears or appears symmetric, reassess as benign or normal (BI-RADS~1 or 2).

\vspace{0.75em}
\textbf{Anti-pattern}\\

\begin{itemize}
\item Labeling a single-view finding as suspicious without multi-view confirmation.
\item Skipping contralateral comparison when suspecting pathology.
\end{itemize}

\end{flushleft}
\end{skillbox}

\begin{skillbox}
\begin{flushleft}

\textbf{Metadata}\\
name: systematic-quadrant-inspection-for-subtle-findings\\
description: When no obvious mass or calcifications are seen, systematically inspect all quadrants using ROI tools to avoid missing subtle asymmetries or microcalcifications.\\
metadata:\\
\hspace*{1em}category: general\\
\hspace*{1em}source: dynamic\\
\hspace*{1em}task: BI-RADS

\vspace{0.75em}
\textbf{Systematic Quadrant Inspection for Subtle Findings}\\

1. Divide the breast into quadrants mentally (upper outer, upper inner, lower outer, lower inner) or use the nipple as reference.\\
2. For each quadrant, use \texttt{inspect\_mammogram\_roi} to zoom in, especially in areas where density appears slightly higher or where the tissue pattern changes.\\
3. Look for clustered calcifications, subtle asymmetries, or architectural distortion that may be obscured by dense tissue.\\
4. Compare the same quadrant in the contralateral breast to decide if a density is focal asymmetry or normal variant.\\
5. Document any finding even if it seems probably benign; do not dismiss as normal without visual confirmation.

\vspace{0.75em}
\textbf{Anti-pattern}\\

\begin{itemize}
\item Stopping after a global impression without quadrant-level inspection.
\item Assuming dense tissue is normal without systematic checking; dense breasts can hide lesions, and systematic inspection reduces undercalling.
\end{itemize}

\end{flushleft}
\end{skillbox}

\begin{skillbox}
\begin{flushleft}

\textbf{Metadata}\\
name: tool-execution-before-conclusion\\
description: After forming a hypothesis about a finding, immediately call the appropriate tool instead of describing a plan. Never present a tool plan without executing it.\\
metadata:\\
\hspace*{1em}category: failure\_recovery\\
\hspace*{1em}source: dynamic\\
\hspace*{1em}task: BI-RADS

\vspace{0.75em}
\textbf{Tool Execution Before Conclusion}\\

1. Formulate the specific question (e.g., ``Is this density visible on the other view?'' or ``What are the margins of this mass?'').\\
2. Immediately call the relevant tool---\texttt{inspect\_mammogram\_roi}, \texttt{inspect\_paired\_mammogram\_view}, or \texttt{inspect\_contralateral\_breast}---without describing the plan in natural language.\\
3. Wait for the tool output before refining your hypothesis. Do not characterize the finding based on the raw image alone.\\
4. After the tool returns, reason from its output. Repeat if necessary but avoid redundant calls.

\vspace{0.75em}
\textbf{Anti-pattern}\\

\begin{itemize}
\item Writing sentences such as ``I will now call inspect\_mammogram\_roi'' without actually invoking the tool. The model must call the tool, not plan to call it.
\end{itemize}

\end{flushleft}
\end{skillbox}

\begin{skillbox}
\begin{flushleft}

\textbf{Metadata}\\
name: calibrate-birads-using-feature-checklist\\
description: Use a systematic feature checklist to avoid extreme BI-RADS assignments (1 or 5) when intermediate categories (2, 3, 4) are more appropriate.\\
metadata:\\
\hspace*{1em}category: general\\
\hspace*{1em}source: dynamic\\
\hspace*{1em}task: BI-RADS

\vspace{0.75em}
\textbf{Calibrate BI-RADS Using Feature Checklist}\\

1. \textbf{BI-RADS~1 (Negative):} No findings on any view. All tool outputs confirm absence of masses, calcifications, asymmetries, or distortion. Both contralateral and paired views show normal symmetric tissue.\\
2. \textbf{BI-RADS~2 (Benign):} Clearly benign findings: popcorn calcifications, vascular calcifications, skin calcifications, well-circumscribed round masses with fat density (e.g., oil cysts, hamartomas).\\
3. \textbf{BI-RADS~3 (Probably Benign):} New, solitary, well-circumscribed solid mass; focal asymmetry that is not changing; grouped punctate calcifications; mild architectural distortion not meeting spiculation criteria. Short-interval follow-up recommended.\\
4. \textbf{BI-RADS~4 (Suspicious):} Indeterminate findings that do not have classic benign features: irregular margins, suspicious calcifications (pleomorphic, linear), new or evolving asymmetry with borderline features.\\
5. \textbf{BI-RADS~5 (Highly Suggestive of Malignancy):} Classic malignant features confirmed on multiple views: spiculated mass, coarse heterogeneous calcifications with linear distribution, architectural distortion with retraction.

\vspace{0.75em}
\textbf{Anti-pattern}\\

\begin{itemize}
\item Jumping from BI-RADS~1 to 5 or 5 to 1 without intermediate evidence.
\item Failing to use the checklist to justify the exact category assigned.
\end{itemize}

\end{flushleft}
\end{skillbox}

\begin{skillbox}
\begin{flushleft}

\textbf{Metadata}\\
name: rule-out-artifact-with-contralateral-comparison\\
description: Before labeling a finding as suspicious, compare with the contralateral breast to identify normal variants that mimic pathology.\\
metadata:\\
\hspace*{1em}category: general\\
\hspace*{1em}source: dynamic\\
\hspace*{1em}task: BI-RADS

\vspace{0.75em}
\textbf{Rule Out Artifact with Contralateral Comparison}\\

1. Invoke \texttt{inspect\_contralateral\_breast} to view the mirror-image region of the opposite breast.\\
2. If a similar density, pattern, or architectural appearance is present in the same location on the opposite side, it is highly likely a normal variant (e.g., asymmetric fibroglandular tissue, inframammary fold).\\
3. If the finding is absent on the contralateral side, then proceed with further characterization.\\
4. Document the comparison result in your reasoning. Bilateral symmetry strongly supports a benign or normal classification (BI-RADS~1 or 2).

\vspace{0.75em}
\textbf{Anti-pattern}\\

\begin{itemize}
\item Concluding a finding is suspicious without checking the opposite breast. Many normal variants are bilateral and symmetric.
\end{itemize}

\end{flushleft}
\end{skillbox}

\begin{skillbox}
\begin{flushleft}

\textbf{Metadata}\\
name: distinguish-benign-from-probably-benign-calcifications\\
description: When evaluating calcifications, differentiate clearly benign patterns (BI-RADS 2) from probably benign patterns (BI-RADS 3) using morphology and distribution.\\
metadata:\\
\hspace*{1em}category: general\\
\hspace*{1em}source: dynamic\\
\hspace*{1em}task: BI-RADS

\vspace{0.75em}
\textbf{Distinguish Benign from Probably Benign Calcifications}\\

1. Use \texttt{inspect\_mammogram\_roi} to zoom in on the calcifications and assess morphology at high resolution.\\
2. \textbf{Benign (BI-RADS~2):} Popcorn (fibroadenoma), coarse (vascular), round/punctate scattered, skin calcifications, milk of calcium, dystrophic, suture. Typically large, well-defined, and not clustered.\\
3. \textbf{Probably Benign (BI-RADS~3):} Grouped fine punctate (5+ in cluster), clustered but monomorphic, small round/oval in a cluster, no pleomorphism or linear shapes. Short-interval follow-up typical.\\
4. \textbf{Suspicious (BI-RADS~4/5):} Pleomorphic, amorphous, fine linear/branching, coarse heterogeneous, with ductal distribution or segmental.\\
5. Always confirm distribution on both views using \texttt{inspect\_paired\_mammogram\_view}.

\vspace{0.75em}
\textbf{Anti-pattern}\\

\begin{itemize}
\item Assigning BI-RADS~2 to a cluster of fine punctate calcifications unless they are clearly scattered and not grouped.
\item Jumping to BI-RADS~4 for grouped punctate calcifications without considering BI-RADS~3.
\end{itemize}

\end{flushleft}
\end{skillbox}

\begin{skillbox}
\begin{flushleft}

\textbf{Metadata}\\
name: avoid-characterizing-findings-without-tool-confirmation\\
description: Never describe a finding as `spiculated', `mass', or `architectural distortion' before using a tool to confirm its appearance.\\
metadata:\\
\hspace*{1em}category: failure\_recovery\\
\hspace*{1em}source: dynamic\\
\hspace*{1em}task: BI-RADS

\vspace{0.75em}
\textbf{Avoid Characterizing Findings Without Tool Confirmation}\\

1. Formulate only as a hypothesis (e.g., ``I see a region of increased density that might be a mass'').\\
2. Immediately call the appropriate tool (\texttt{inspect\_mammogram\_roi}, \texttt{inspect\_paired\_mammogram\_view}, or \texttt{inspect\_contralateral\_breast}) to gather evidence.\\
3. After receiving tool output, use the observed features to characterize the finding. Describe what the tool shows, not what you think you see in the raw image.\\
4. If the tool output does not clearly show the feature, do not assert it in reasoning.

\vspace{0.75em}
\textbf{Anti-pattern}\\

\begin{itemize}
\item Stating ``The mammogram reveals a spiculated mass'' before any tool call. Such statements are hallucinations and lead to misclassification.
\end{itemize}

\end{flushleft}
\end{skillbox}

\begin{skillbox}
\begin{flushleft}

\textbf{Metadata}\\
name: structured-reasoning-grounded-in-tool-outputs\\
description: Ensure every statement in the reasoning summary is directly supported by a tool finding listed in tool\_findings, with explicit citation.\\
metadata:\\
\hspace*{1em}category: failure\_recovery\\
\hspace*{1em}source: dynamic\\
\hspace*{1em}task: BI-RADS

\vspace{0.75em}
\textbf{Structured Reasoning Grounded in Tool Outputs}\\

1. List every tool call in \texttt{tool\_findings} in the exact order they were made, with one bullet per tool.\\
2. In \texttt{reasoning\_summary}, for each claim about the image, explicitly reference which tool provided the evidence (e.g., ``\texttt{inspect\_paired\_mammogram\_view} confirmed the finding is present on both views'').\\
3. Do not include any visual observations that were not obtained from a tool output. If an observation was made from the raw image and then confirmed with a tool, only the tool output counts as evidence.\\
4. If multiple tools were used, explain how each tool contributed to the final decision.

\vspace{0.75em}
\textbf{Anti-pattern}\\

\begin{itemize}
\item Writing a reasoning summary that contains descriptions of findings without attributing them to specific tool calls. Every piece of evidence must be traceable to a tool output.
\end{itemize}

\end{flushleft}
\end{skillbox}

\begin{skillbox}
\begin{flushleft}

\textbf{Metadata}\\
name: use-uncertainty-resolution-sequence\\
description: When uncertain about a finding, follow a structured sequence: inspect ROI, check paired view, compare contralateral, and optionally retrieve knowledge before finalizing.\\
metadata:\\
\hspace*{1em}category: general\\
\hspace*{1em}source: dynamic\\
\hspace*{1em}task: BI-RADS

\vspace{0.75em}
\textbf{Use Uncertainty Resolution Sequence}\\

1. \textbf{Step 1:} Use \texttt{inspect\_mammogram\_roi} (up to 2 times) to zoom in on the area and characterize margins, shape, density, and internal features.\\
2. \textbf{Step 2:} Use \texttt{inspect\_paired\_mammogram\_view} to confirm the finding appears in the second projection. If it does not, it is likely superimposition.\\
3. \textbf{Step 3:} Use \texttt{inspect\_contralateral\_breast} to check if the finding is bilateral. If symmetric, it is a normal variant.\\
4. \textbf{Step 4 (if needed):} Use \texttt{retrieve\_knowledge} to resolve definitional uncertainty about categories (e.g., what exactly constitutes a probably benign calcification).\\
5. Only after completing this sequence should you assign a BI-RADS category.

\vspace{0.75em}
\textbf{Anti-pattern}\\

\begin{itemize}
\item Making a final decision while still uncertain. The sequence is designed to reduce uncertainty incrementally.
\item Skipping steps, which leads to over- or under-calling.
\end{itemize}

\end{flushleft}
\end{skillbox}

\end{document}